\input{aipcheck}

\documentclass[
    ,final            % use final for the camera ready runs
  ]
  {aipproc}

\layoutstyle{6x9}
\newcommand{\beq}{\begin{equation}}
\newcommand{\eeq}[1]{\label{#1}\end{equation}}
\newcommand{\beqa}{\begin{eqnarray}}
\newcommand{\eeqa}[1]{\label{#1}\end{eqnarray}}
\newcommand{\eeqan}{\end{eqnarray}}

\makeatletter
\def\compoundrel#1\over#2{\mathpalette\compoundreL{{#1}\over{#2}}}
\def\compoundreL#1#2{\compoundREL#1#2}
\def\compoundREL#1#2\over#3{\mathrel
      {\vcenter{\hbox{$\m@th\buildrel{#1#2}\over{#1#3}$}}}}
\makeatother

\begin{document}

\title{Chiral model for $\bar{K}N$ interactions\\
and its pole content}


\author{A. Ciepl\'{y}}{
  address={Nuclear Physics Institute, 250 68 \v{R}e\v{z}, Czech Republic}
}

\author{J. Smejkal}{
  address={Institute of Experimental and Applied Physics, 
Czech Technical University in Prague, Horsk\'{a}~3a/22, 128~00~Praha~2, 
Czech Republic}
}

\begin{abstract}
We use chirally motivated effective meson-baryon potentials to describe the low energy 
$\bar{K}N$ data including the characteristics of kaonic hydrogen. Our results are 
examined in comparison with other approaches based on the unitarity and dispersion 
relation for the inverse of the T-matrix. We demonstrate that the movements of the poles 
generated by the model upon varying the model parameters can serve as a tool to get 
additional insights on the dynamics of the strongly coupled $\pi\Sigma$-$\bar{K}N$ system.
\end{abstract}

\maketitle

%%%%%%%%%%%%%%%%%%%%%%%%%%%%%%%%%%%%%%%%%%%%
%% MAINMATTER
%%%%%%%%%%%%%%%%%%%%%%%%%%%%%%%%%%%%%%%%%%%%

\section{Introduction}

The synergy of chiral perturbation theory and the coupled channel T-matrix resummation 
techniques has proven to provide a successful description of $\bar{K}N$ interactions 
at low energies. Although some issues still remain to be resolved (e.g. the compatibility 
of $K^{-}p$ scattering and kaonic hydrogen data or the nature of the $\Lambda(1405)$ resonance) 
there is a hope that the coming experimental data, particularly those from SIDDHARTA 
collaboration, will shred more light on the topic. In our report we briefly examine and compare 
the available theoretical models while paying an attention to their pole content and to their
predictions for the kaonic hydrogen atom characteristics. We also demonstrate the importance 
of the so called zero coupling limit which allows to relate the poles (of the coupled 
channel S-matrix, or of the $\bar{K}N$ amplitude) found on the complex energy manifold 
to the meson-baryon channels. Interestingly, we find that the chirally motivated effective 
potentials generate not only two isoscalar poles related to the $\Lambda(1405)$ resonance 
but two isovector poles as well.

In our approach we employ chirally motivated coupled-channel potentials that are taken
in a separable form,
\beq
V_{ij}(k,k')=\sqrt{\frac{1}{2E_i}\frac{M_i}{\omega_i}} g_{i}(k^{2})
             \; \frac{C_{ij}}{f_{\pi}^2} \;
             g_{j}(k'^{2}) \sqrt{\frac{1}{2E_j}\frac{M_j}{\omega_j}} \;\; ,
 \;\;\;\;\; g_{j}(k)=\frac{1}{1+(k/ \alpha_{j})^2} \; ,
\eeq{eq:Vpot}
with $E_i$, $M_i$ and $\omega_i$ denoting the meson energy, the baryon mass
and baryon energy in the c.m. system of channel $i$. The coupling
matrix $C_{ij}$ is determined by the chiral SU(3) symmetry. 
The parameter $f_{\pi} \sim 100$ MeV represents the pseudoscalar meson decay constant 
in the chiral limit, and the inverse range parameters $\alpha_{i}$ are fitted 
to the low energy $\bar{K}N$ data. The indexes $i$ and $j$ run over the set of 
coupled meson-baryon channels composed from the $\pi\Lambda$, $\pi\Sigma$, $\bar{K}N$, 
$\eta\Lambda$, $\eta\Sigma$, and $K\Xi$ states (taken with all appropriate charge 
combinations). The details of our model are given in Ref.~\cite{10CSm}.

The chiral symmetry of meson-baryon interactions is reflected 
in the structure of the $C_{ij}$ coefficients derived directly from the 
Lagrangian. An exact content of the coefficients up to second order in the meson 
c.m. kinetic energies was specified in Refs.~\cite{95KSW} and \cite{10CSm}. 
In practice, many authors (e.g. \cite{03JOO}, \cite{08HWe}) consider only the leading 
order Weinberg-Tomozawa interaction with the energy dependence defined by
\beq
C_{ij} = - C_{i j}^{\rm (WT)} (2\sqrt{s} -M_{i} -M_{j})/4 \;\;\; .
\eeq{eq:CWT}
One should note that this relativistic prescription differs from the one adopted 
in models derived from the chiral Lagrangian formulated for static baryons \cite{10CSm}, 
\cite{95KSW} and expanded strictly only to the second order in meson energies and quark masses. 
In principle, the approaches based on different formulations of the chiral 
Lagrangian should give the same results for physical observables. However, 
this is true only when one sums up all orders of the infinite series 
of the relevant Feynman diagrams (all orders in $q$), not once we restrict 
ourselves to a given perturbative order. In other words, the models based on 
various Lagrangian formulations or models that vary in their prescriptions 
for treatment of the terms beyond the leading order may give (to a reasonable extent) 
different predictions for the measurable quantities.

\section{Data reproduction and model comparison}

The available experimental data on low energy $\bar{K}N$ interactions consist of
\begin{itemize}
\item the $K^- p$ cross sections for the elastic scattering and reactions to the $\bar{K^0}n$, 
$\pi^{+} \Sigma^{-}$, $\pi^{-} \Sigma^{+}$, $\pi^0 \Lambda$, and $\pi^{0} \Sigma^{0}$ channels 
(see references collected in \cite{95KSW})
\item the $K^- p$ threshold branching ratios, standardly denoted as $\gamma$, $R_c$, and $R_n$ 
\cite{81Mar}
\item the kaonic hydrogen characteristics, the strong interaction shift of the $1s$ energy level 
$\Delta E_{N}$ and the decay width of the $1s$ level $\Gamma$ \cite{98KEK}, \cite{05DEAR}
\end{itemize}
In general, the chirally motivated models have no problem with reproduction of the low energy 
$K^- p$ cross sections, mostly due to relatively large error bars of the experimental data. 
The threshold branching ratios are determined with much better precision and provide a sterner 
test for any quantitative usage of the models. The kaonic hydrogen data are represented 
by an older (and rather unprecise) KEK measurement \cite{98KEK} and a more recent 
DEAR measurement \cite{05DEAR}. Although the later experiment determined the $K^{-}$-atomic
characteristics with sufficiently good resolution 
it was found to be at odds with the scattering data extrapolated to the $K^{-}p$ threshold. 
The situation should be resolved soon thanks to the already completed SIDDHARTA experiment 
the data of which are analyzed.
In addition, one should also consider the $\pi\Sigma$ mass distribution generated by the models 
and compare it with the data that reveal a resonance just below the $\bar{K}N$ threshold. 
The observed mass distribution is assigned to the isoscalar $\Lambda(1405)$ resonance. The chiral 
models provide two isoscalar resonances that overlap in the appropriate energy region. Though, 
it is still not so well determined what are the positions of the pertinent poles 
in the complex energy plane 
and whether both of them are sufficiently close to the real axis to affect physical observables.

In the Table \ref{tab:models} we compare the predictions various models provide 
for the 1s level characteristics of kaonic hydrogen, for the branching ratios at the $K^{-}p$
threshold and for the positions $z_1$ and $z_2$ of the poles related to $\Lambda(1405)$ 
on the second Riemann sheet. The first three rows 
are represented by models that include only the leading Weinberg-Tomozawa (WT) interaction, 
i.e. the interchannel couplings comply with Eq.~(\ref{eq:CWT}). The model WT1 represents our 
own fit (to the kaonic hydrogen KEK data plus the $K^{-}p$ cross sections and branching ratios) 
with the pion decay constant fixed at $f_{\pi}$ = 107 MeV and with the common inverse range 
$\alpha$ used as the only free parameter in the fit (we got $\alpha$ = 650 MeV). The other 
two WT models are taken from Refs.~\cite{03JOO} and \cite{08HWe}. The next two lines 
represent models that include the NLO corrections to the interchannel couplings $C_{ij}$ 
and the last line in Table \ref{tab:models} shows the experimental data (the pole positions 
are not measurable quantities). The models CS30 and BNW (we picked only one representative 
from those available in the respective papers) used the DEAR data (rather then the less 
precise KEK values) to fit the kaonic hydrogen characteristics. 
Apparently, one can achieve almost perfect reproduction of the experimental branching ratios 
(though some models did not aim at it) but the computed kaonic hydrogen characteristics 
are off the data reported in the DEAR experiment. This applies specifically to the 1s level 
decay width that all models predict about three standard deviations above the measured value.

\begin{table}
\begin{tabular}{c|cc|ccc|cc}
  model & $\Delta E_{N}$ [eV] & $\Gamma$ [eV]& $\gamma$ & $R_c$ & $R_n$ & $z_1$ [MeV] & $z_2$ [MeV] 
                                                                                             \\ \hline
 WT1                & 366       & 696       & 2.366   & 0.636 & 0.188 & (1360,-54)  & (1431,-21) \\
 HW \cite{08HWe}    & 270$^{*}$ & 570$^{*}$ & 1.80    & 0.624 & 0.225 & (1400,-76) & (1428,-17)  \\
 JOORM \cite{03JOO} & 275$^{*}$ & 586$^{*}$ & 2.30 & 0.618 & 0.257 & (1389,-64) & (1427,-17) \\ \hline
 CS30 \cite{10CSm} & 260     & 692         & 2.366   & 0.655 & 0.188 & (1398,-51) & (1441,-76) \\ 
 BNW \cite{05BNW}  & 236$^{*}$ & 580$^{*}$ & 2.35 & 0.653 & 0.194 & (1408,-37) & (1449,-106) \\ \hline
 exp               & 193(43) & 249(150) & 2.36(4) & 0.664(11) & 0.189(15) & -- & -- \\
\end{tabular}
\caption{Model predictions for the $\bar{K}N$ threshold data. The values $\Delta E_{N}$
and $\Gamma$ marked by stars were established from the $K^{-}p$ scattering length by means of the modified Deser-Trueman relation \cite{04MRR}.}
\label{tab:models}
\end{table}

It is remarkable that all WT models more or less agree on the position of the pole $z_2$. However,
this agreement is spoiled (and the pole does not appear so close to the real axis) when the NLO 
corrections are included in the interchannel couplings. Our understanding is that the parameter 
space becomes too large when the second order couplings (low energy constants) have to be fitted 
and then the experimental data allow for more local minimums of the $\chi^{2}$, each of them 
leading to a different position of the pole. The position of the pole $z_1$ varies depending on 
the particular model even in the case of interaction reduced purely to the leading Weinberg-Tomozawa 
coupling. One can only say that it is located much further from the real axis then the pole $z_2$ 
and lies at lower energies (in terms of Re$\;z$) than $z_2$. The pole $z_2$ is normally identified 
as the one that relates to the subthreshold behavior of the $K^{-}p$ amplitude and to 
the $\Lambda(1405)$ resonance observed in the $\pi\Sigma$ mass spectrum in $\bar{K}N$ 
initialized reactions. However, since the NLO corrections seem to push the $z_2$ pole above 
the $K^{-}p$ threshold and further from the real axis it may be the $z_1$ pole that affects 
significantly the physical observables below the $\bar{K}N$ threshold. Thus we clearly need 
more experimental data that would clarify the role of the NLO corrections and the nature 
of the $\Lambda(1405)$ resonance as such.

\section{Poles origin and their movements}

\begin{figure}
  \includegraphics[height=.4\textheight]{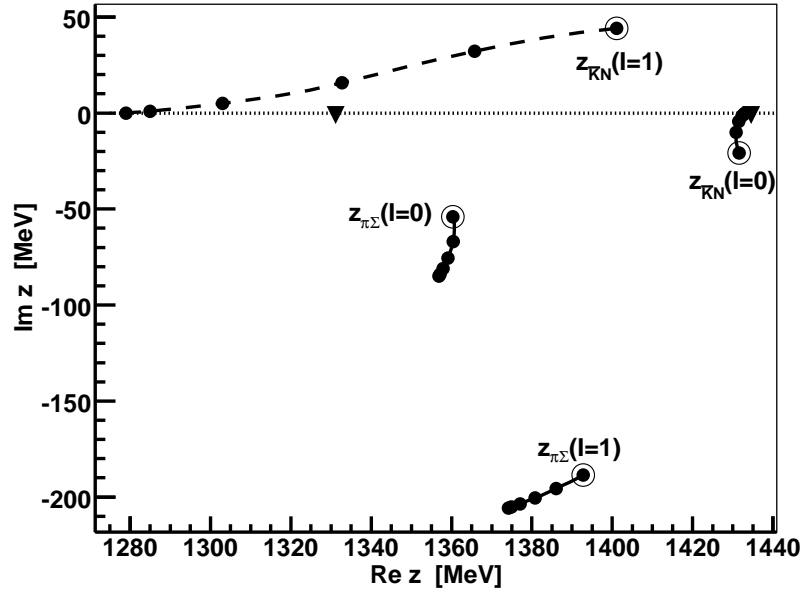}
  \caption{Pole movements upon scaling the nondiagonal interchannel couplings. The bullets 
   visualize the pole positions for $x=0$, $0.2$, ... , $1$. The "physical" positions 
   for $x=1$ are also encircled and the $\pi\Sigma$ and $\bar{K}N$ thresholds 
   are marked by triangles.}
  \label{fig:poles}
\end{figure}

The positions of the poles of the scattering S-matrix (or of the appropriate T-matrix) 
affect physical observables provided the poles are close to the physical region. When the parameters
of the model are varied the poles move on the complex energy manifold and can even move 
from one Riemann sheet (RS) to another one by crossing the real axis. It is instructive to look 
at the pole movements when one gradually reduces the interchannel couplings while keeping 
the diagonal couplings intact. One can do so by multiplying the nondiagonal 
couplings $C_{ij}$, $i \neq j$, by a scaling parameter $x$ with $x=1$ standing for the physical 
limit and $x=0$ for the so called zero coupling limit. For $x=0$ the positions of the poles 
can be found as solutions of a simple equation that relates the diagonal coupling 
(or the separable potential $V_{ij}$, $i=j=n$) to the Green function $G_{n}$,
\beq
\frac{4\pi f_{\pi}^{2}}{C_{nn}}\: \frac{z}{M_n} + G_{n}(z) = 0 \;\;\; ,
\eeq{eq:x0}
where the complex energy $z$ is equal to the meson-baryon cms energy $\sqrt{s}$ 
at the real axis. In Figure \ref{fig:poles} we show the pole trajectories as they evolve 
from the zero coupling limit to their physical locations. The trajectories were computed 
for our WT1 model separately for the isoscalar channels ($I=0$ states of $\pi \Sigma$, 
$\bar{K}N$, $\eta \Lambda$, $K\Xi$) and for the isovector ones ($I=1$ states of $\pi \Lambda$, 
$\pi \Sigma$, $\bar{K}N$, $\eta \Sigma$, $K\Xi$). The trajectories of both the isoscalar 
as well as the isovector poles are presented in one figure with the [+,-] RS 
shown for Im$\;z\;>\:0$ and the [-,+] RS for Im$\;z\;<\:0$. In our notation 
the first and the second $+/-$ refer to the signs of the imaginary part of the relative 
$\pi \Sigma$ and $\bar{K}N$ momenta, respectively (the [-,+] RS is standardly referred to 
as the second RS reached by crossing the real axis in between the $\pi \Sigma$ and $\bar{K}N$ 
thresholds). A similar analysis can be done with 
the physical channels, though then the interpretation is not so straightforward as the scaling 
of interchannel couplings breaks the isospin symmetry and it is not possible to define proper 
isospin states for $x<1$.

The zero coupling limit enables us to assign the poles unambiguously either to the $\pi\Sigma$ 
or to the $\bar{K}N$ states while they couple to both channels for $x>0$.
We get two isoscalar poles, one related to the $\pi \Sigma$ 
channel and another one to the $\bar{K}N$ channel. They correspond to the solutions 
of Eq.~(\ref{eq:x0}) for the pertinent channels. Both poles develop on the [-,+] RS with 
the $\pi \Sigma$ starting (for $x=0$) as a resonant pole [$z = (1357 - {\rm i}\:85)$ MeV] 
and the $\bar{K}N$ pole as a bound state pole ($z = 1433$ MeV), just below the $\bar{K}N$ 
threshold. Interestingly, we also find two isovector poles. The one related to $\pi \Sigma$ 
is on the same [-,+] RS as the isoscalar poles and it is located very far from the 
real axis [$z = (1393 - {\rm i}\:189)$ MeV for $x=1$]. The $\bar{K}N$ isovector pole starts its 
movement in the zero coupling limit as a virtual $\bar{K}N$ state and then develops 
on the [+,-] RS. Its position in the physical limit (for $x=1$) is $z = (1401 - {\rm i}\:33)$ MeV. 
Although, it appears on a RS that is not directly connected with the physical region, 
it is relatively close to the $\bar{K}N$ threshold and to the real axis, so it does affect 
the threshold behavior of the elastic $K^{-}n$ amplitude. The two isoscalar poles are those 
that are standardly assigned to the $\Lambda(1405)$ resonance and their positions in 
the physical limit are those given it the Table \ref{tab:models} for our WT1 model. 
The isovector $\bar{K}N$ pole was already discussed in Ref.~\cite{03JOO}. As far as we know, 
we are the first to report on the isovector $\pi\Sigma$ pole. Though, it lies 
very far from the real axis and hardly affects any physical observables its existence 
may be important in view of the SU(3) symmetry as well as the pole content of the chiral models 
discussed here and elsewhere.

%%%%%%%%%%%%%%%%%%%%%%%%%%%%%%%%%%%%%%%%%%%%%%%%
%% BACKMATTER
%%%%%%%%%%%%%%%%%%%%%%%%%%%%%%%%%%%%%%%%%%%%%%%%

\begin{theacknowledgments}
We are grateful to A.Gal for stimulating our work on the pole movements. The work is supported 
by the Grant Agency of the Czech Republic, grant No.\ 202/09/1441.
\end{theacknowledgments}



\bibliographystyle{aipproc}   % if natbib is available

\begin{thebibliography}{99}

\bibitem{10CSm} 
A.~Ciepl\'{y} and J.~Smejkal, Eur.~Phys.~J.~A 43 (2010) 191.

\bibitem{95KSW} N.~Kaiser, P.B.~Siegel, and W.~Weise, Nucl.~Phys.~A 594 (1995) 325.

\bibitem{03JOO} D. Jido, J.A. Oller, E. Oset, A. Ramos and U.-G. Meissner, 
  Nucl.\ Phys.\ A 725 (2003) 181.

\bibitem{08HWe} 
T.~Hyodo and W.~Weise, Phys.\ Rev.\ C 77 (2008) 035204.

\bibitem{81Mar} A.D. Martin, Nucl.~Phys.~B 179 (1981) 33; and earlier
references cited therein.

\bibitem{98KEK} M.~Iwasaki {\it et al.}, Phys.\ Rev.\ Lett.\ 78 (1997) 3067; 
              T.M.~Ito {\it et al.}, Phys.\ Rev.\ C 58 (1998) 2366.

\bibitem{05DEAR}  G. Beer {\it et al}.\ [DEAR Collab.], 
Phys.\ Rev.\ Lett.\ 94 (2005) 212302.

\bibitem{05BNW} B.~Borasoy, R.~Nissler, and W.~Weise, 
Eur.~Phys.~J.~A 25 (2005) 79.

\bibitem{04MRR} 
U.-G.~Meissner, U.~Raha, and A.~Rusetsky, Eur.~Phys.~J.~C 35 (2004) 349.


\end{thebibliography}

\end{document}